\documentstyle[11pt]{article}
\newcommand{\DS}{/\!\!\!\! D}
\renewcommand{\bar}{\overline}
\begin{document}
\begin{flushright}
SNUTP-97-038 \\
hep-th/9703185\\
\today
\end{flushright}
\begin{center}
{\LARGE $N=2$ Supersymmetric Gauged $O(3)$ Sigma Model}\\[4mm]
{\large Kyoungtae Kimm and  Bukyung Sul}\\[2mm]
Center for Theoretical Physics, Seoul National University, \\
Seoul 151-742, Korea\\
\end{center}
\centerline{\large\bf Abstract}
\begin{quote}
We discuss the $N=2$ supersymmetric extension of the gauged $O(3)$
sigma model in $(2+1)$ dimensions with an abelian Chern-Simons term.
It is shown that the self-dual potential and the Bogomolny relations
naturally appear as consequences of extended supersymmetry.
\end{quote}
\vskip 2em

Recently there have been several works in the gauged $O(3)$ sigma model
in (2+1) dimensions with a Chern-Simons term where the gauge group is
$SO(2)$ ( or $U(1)$ ) subgroup of $O(3)$~\cite{kimm}.
It was found that in this Chern-Simons $O(3)$ sigma model 
the energy functional obeys the Bogomolny-type lower bound~\cite{bogo} 
when a specific potential is chosen.
The bound is saturated by the Bogomolny equations (or self-dual equations)
which are the first order differential equations and solve the equations of
motion automatically.
This system in some respects has similarity  to the 
Chern-Simons Higgs model where 
the Bogomolny relations are also 
achieved with a specific potential~\cite{hong}, but there are much richer
soliton spectrums in this model.

The specific form of potential in the Chern-Simons Higgs model was guaranteed
by introducing the $N=2$ supersymmetry~\cite{clee} 
and this supersymmetry can be expected
for every self-dual model~\cite{witten78}.
In the $N=2$ supersymmetric extension of self-dual models the topological
charge which constitutes the Bogomolny bound 
appears a central charge of supersymmetry algebra, and from that algebra
the Bogomolny relations are obtained.

In this paper we explicitly show 
that the Chern-Simons $O(3)$ sigma model also
admits $N=2$ supersymmetric extension. To be specific, we construct
the $N=1$ supersymmetric version of our model, and then 
show that requirement of 
$N=2$ supersymmetry  determines superpotential.
Finally, we obtain the Bogomolny relations and the topological charge from 
the supersymmetry algebra.

Let us begin with the model given by the action 
\begin{eqnarray}
S=\int d^3x \Big\{
\frac{1}{2} (D_\mu \phi^a)^2 
+\frac{1}{2}\kappa\epsilon^{\mu\nu\rho} A_\mu\partial_\nu A_\rho
- U(n^a\phi^a)\Big\}.
\label{action1}
\end{eqnarray}
Here $\phi^a$ $(a=1,2,3)$ is an $O(3)$ sigma field 
with the constraint $\phi^a\phi^a=1$.
We gauge the $SO(2)$ rotational symmetry along a fixed axis $n^a$ in 
the internal space of $O(3)$ sigma field $\phi^a$. 
The gauge interaction is introduced through the covariant derivative 
$D_\mu\phi^a=\partial_\mu\phi^a+A_\mu(n\times\phi)^a$ where 
$(n\times\phi)^a=\epsilon^{abc}n^b\phi^c$.
When we choose the following form of  potential~\cite{kimm} 
\begin{eqnarray}
\label{pot}
U=\frac{1}{2\kappa^2}(v-n^a\phi^a)^2(n\times \phi)^2,
\end{eqnarray}
with a real parameter $v$, 
it can be shown that
this model develops the Bogomolny-type bound in energy functional
\begin{eqnarray}
E&=& \int d^2x \Big\{ \frac{1}{2}(D_\mu\phi^a)^2 
      +\frac{1}{2\kappa^2}(v-n^a\phi^a)^2(n\times\phi)^2\Big\} \nonumber\\
 &=&\int d^2x \Big\{ 
\frac{1}{2}\Big(D_0\phi^a \pm 
      \frac{1}{\kappa}(v-n^b\phi^b)(n\times\phi)^a\Big)^2
    +\frac{1}{4}\Big(D_i\phi^a
\pm \epsilon_{ij} (\phi\times D_j \phi)^a\Big)^2\Big\}
    \pm T           \label{energy} \\
&\ge& |T|\nonumber ,
\end{eqnarray}
where $T$ is the spatial integral of temporal component of 
topological current $K^\mu=\frac{1}{2}\epsilon^{\mu\nu\rho}(\phi^a
(D_\nu \phi\times D_\rho\phi)^a+(v-n^a\phi^a) F_{\nu\rho})$, which is
gauge-invariant and conserved. 
The energy bound in Eq.~(\ref{energy}) is saturated 
by the following Bogomolny equations
\begin{eqnarray}
\label{bog1}
D_i\phi^a \pm\epsilon_{ij}(\phi\times D_j \phi)^a=0, \\
\epsilon^{ij}F_{ij} 
\mp\frac{2}{\kappa^2}(v-n^a\phi^a)(n\times\phi)^2=0,
\label{bog2}
\end{eqnarray} 
where we have used the Gauss law constraint 
$\frac{1}{2}\kappa \epsilon^{ij} F_{ij} =-(n\times\phi)^aD_0 \phi^a$
that follows from the action
(\ref{action1}) under the variation with respect to $A_0$.
Depending on the value of $v$ there are symmetric or asymmetric phases and 
various self-dual
solitons -- lumps, vortices and nontopological solitons -- of this model
whose properties are now well-understood.

In order to make supersymmetric extension of this model we  need a
three-component scalar superfield $\Phi^a$ which reads in the superspace
\begin{eqnarray}
\Phi^a=\phi^a +\bar\theta\psi^a 
+\frac{1}{2}\bar\theta\theta F^a,
\label{sfield}
\end{eqnarray}
where $\phi^a$ is scalar field and $\psi^a$ is Majorana spinor,
and $F^a$ is auxiliary field. 
$\theta_\alpha$ is a two-component Majorana spinor. 
We use the convention for $\gamma$-matrices  as
$\gamma^0=\sigma_2$, $\gamma^1=i\sigma_3$, $\gamma^2=i\sigma_1$
and  define $C_{\alpha\beta}=(\sigma_2)=-C^{\alpha\beta}$. 
A Majorana spinor with lower index is then real in this representation. 
The spinors with upper index carry an upperbar for convenience, i.e.,  
$\bar\theta{}^\alpha=C^{\alpha\beta}\theta_\beta$, and contraction of
spinor index means 
$\bar\theta\theta
=C^{\alpha\beta}\theta_\beta\theta_\alpha
=\bar\theta{}^\alpha\theta_\alpha$~\cite{gates}.
The superfield (\ref{sfield}) of sigma model  is  constrained 
to satisfy $\Phi^a \Phi^a=1$,
which yields three constraints 
\begin{eqnarray}
\label{const}
\phi^a\phi^a &=&1,\nonumber\\
\phi^a\psi^a&=&0 ,\\ 
\phi^a F^a&=&\frac{1}{2}\bar\psi{}^a\psi^a, \nonumber 
\end{eqnarray}
when expanded in powers of $\theta$.
In addition, we introduce a real spinorial superfield $W_\alpha$ which in the 
Wess-Zumino gauge contains the Chern-Simons gauge field 
$A_\mu$ and the gaugino $\lambda_\alpha$.

With these the $N=1$ supersymmetric 
extension of model in Eq.~(\ref{action1}) is given by a superspace action 
\begin{eqnarray}
S&=&\int d^3xd^2\theta \Big\{-\frac{1}{4}
( D^\alpha \Phi^a + W^\alpha (n\times\Phi)^a) 
(D_\alpha \Phi^a +W_\alpha (n\times\Phi)^a)  
+f(n^a\Phi^a)\nonumber \\ 
&&-\frac{1}{8}\kappa D^\alpha W^\beta D_\beta W_\alpha
\Big\}.
\label{action2}
\end{eqnarray}
The superpotential $f$ is the function of $n^a\Phi^a$ alone in order to
preserve the gauge symmetry, and will explicitly be specified  in the below.

With the integration over $\theta$ the action takes the form in components
\begin{eqnarray}
S&=&\int d^3x\Big\{\frac{1}{2}(D_\mu \phi^a)^2 
+\frac{1}{2}i\bar\psi{}^a\DS\psi^a
+\frac{1}{2} (F^a)^2 
+\frac{1}{2}\kappa \epsilon^{\mu\nu\rho} A_\mu \partial_\nu A_\rho
-\frac{1}{2}\kappa \bar\lambda\lambda \nonumber \\
&&-\bar\lambda\psi^a(n\times\phi)^a 
-(F^a n^a) f'
+\frac{1}{2}(n^a\bar\psi{}^a)(\psi^bn^b)f''\Big\}
\label{action3},
\end{eqnarray}
where $f'=f'(n^a\phi^a)$ and $f''=f''(n^a\phi^a)$.
It can easily be checked that the action (\ref{action3})  
is invariant under the
following supersymmetry transformations:
\begin{eqnarray}\label{susy1}
\delta \phi^a&=&-\bar\eta \psi^a ,\nonumber\\
\delta \psi^a&=&-\eta F^a
 + i \DS\phi^a\eta,  \nonumber\\
\delta F^a &=&i\bar\eta~\DS\psi^a
            -2\bar\eta\lambda (n\times\phi)^a, \\
\delta A^\mu&=&-i \bar\eta\gamma^\mu\lambda,\nonumber\\
\delta \lambda
&=&i\epsilon_{\mu\nu\sigma}
       \partial^\nu A^\sigma\gamma^\mu\eta \nonumber ,
\end{eqnarray}
where $\eta_\alpha$ is an infinitesimal Majorana spinor.
The conserved supercurrent associated with these transformations (\ref{susy1}) 
is 
$J_{N=1}^\mu =D_\nu\phi^a \gamma^\nu\gamma^\mu \psi^a
             -i\gamma^\mu\psi^a n^a f' $, 
and the corresponding supercharge is 
\begin{eqnarray}
\label{charge1}
Q^{(1)}=\int d^2x \{ D_\nu\phi^a\gamma^\nu \gamma^0 \psi^a
                     -i \gamma^0\psi^a n^a f'\}.
\end{eqnarray}

The auxiliary fields $F^a$ and $\lambda$ can be removed from the action 
by solving their equations of motion:
\begin{eqnarray}
\lambda&=&-\frac{1}{\kappa} \psi^a(n\times\phi)^a ,\\
F^a&=& (n^a -\phi^a(\phi^bn^b) )f' 
+\frac{1}{2}(\bar\psi{}^b\psi^b)\phi^a,
\end{eqnarray}
where we have taken the constraints (\ref{const}) 
into account by using Lagrange multipliers.
Inserting these into the action (\ref{action3}), 
we have:
\begin{eqnarray}
S&=& 
\int d^3x\Big\{\frac{1}{2}(D_\mu\phi^a)^2
+\frac{1}{2}i\bar\psi{}^a \DS\psi^a
-\frac{1}{2}(f')^2(n\times\phi)^2
+\frac{1}{2}\kappa \epsilon^{\mu\nu\rho}A_\mu\partial_\nu A_\rho
 \nonumber \\
&&-\frac{1}{2}\Big(f'-\frac{1}{\kappa}(n\times\phi)^2\Big)(\bar\psi{}^a\psi^a)
+\frac{1}{2}\Big( f''-\frac{1}{\kappa}\Big)(n^a\bar\psi{}^a)(\psi^bn^b)
+\frac{1}{8}(\bar\psi{}^a\psi^a)^2 
\Big\}.
\label{action4}
\end{eqnarray}

{}From the form of action (\ref{action4}) and the constraint (\ref{const})
one can easily notice that there
is an additional $O(2)$ symmetry in this  model 
if we restrict the superpotential further,
which is necessary to require $N=2$ extended supersymmetry.
To be specific, if we impose 
\begin{eqnarray}
f''=\frac{1}{\kappa}, 
\end{eqnarray}
on the superpotential we have the following symmetry
$\delta\psi^a=\varepsilon (\phi\times\psi)^a$.
It is a local symmetry, however, the action is invariant
under this transformation
in view of the fact that $\phi^a$ and $\psi^a$ are mutually orthogonal.
The corresponding conserved Noether charge is 
$C=\int d^2x(\phi\times\bar\psi )^a\gamma^0\psi^a$~\cite{witten77}.
Now the superpotential can be written as 
\begin{eqnarray}
\label{sp}
f(n^a\Phi^a)=\frac{1}{2\kappa}(v-n^a\Phi^a)^2,
\end{eqnarray}
where $v$ is the parameter of the model. As expected, the bosonic sector of 
the supersymmetric action (\ref{action4}) 
with superpotential (\ref{sp}) 
reproduces the action of the self-dual gauged $O(3)$ 
sigma model in Eq.~(\ref{action1}).

With the help of charge $C$, we can find the second conserved
supercurrent 
$J^{\mu}_{N=2}=(\phi\times D_\nu\phi)^a\gamma^\nu\gamma^\mu\psi^a
+i(n\times\phi)^a\gamma^\mu \psi^a f'$ and its supercharge
\begin{eqnarray}
\label{charge2}
Q^{(2)} =\int d^2x \Big\{
 (\phi\times D_\nu\phi)^a \gamma^\nu\gamma^0\psi^a
+i(n\times\phi)^a \gamma^0 \psi^a f'\Big\} .
\end{eqnarray}
The supercharges $Q^{(1)}_\alpha$, $ Q^{(2)}_\alpha$ 
and $C$ generate the algebra
which can be understood as the supersymmetry transformations with internal
$O(2)$ symmetry:
\begin{eqnarray}
\big[C, Q^{(1)}_\alpha \big] &=& Q^{(2)}_\alpha ,\nonumber \\
\big[C, Q^{(2)}_\alpha \big] &=&-Q^{(1)}_\alpha ,\nonumber
\end{eqnarray}
and
\begin{eqnarray}
\label{alg}
\{ Q^{(A)}_\alpha ,Q^{(B)}_\beta \}
=2P_\mu (\gamma^\mu)_{\alpha\beta}\delta^{AB}
+ 2Z\epsilon_{\alpha\beta}\epsilon^{AB}.
\end{eqnarray}
Here $Z$ is the central charge defined by
\begin{eqnarray}
\label{cent1}
Z= \int d^2x \Big\{\epsilon^{ij}\phi^a (D_i\phi\times D_j\phi)^a
+\Big(2 (n\times \phi)^a D_0\phi^a 
-i (n\times\bar\psi)^a \gamma^0 \psi^a \Big) f'\Big\}.
\end{eqnarray}
With the Gauss law constraint which  
follows from the variation of action
(\ref{action4}) with respect to $A_0$  we have 
$\kappa\epsilon^{ij} F_{ij} =-2(n\times\phi)^a D_0\phi^a
+i (n\times\bar\psi)^a \gamma^0 \psi^a $, so that we get
\begin{eqnarray}
\label{cent2}
Z= \int d^2x \epsilon^{ij}\Big\{\phi^a (D_i\phi\times D_j\phi)^a
+(v-n^a\phi^a)F_{ij}
\Big\}.
\end{eqnarray}
The central charge $Z$  coincides exactly with the 
topological charge $T$ of the model (\ref{action1}) with 
self-dual potential (\ref{pot}). 

We are interested in providing the connection between 
the supercharge algebra (\ref{alg}) and 
Bogomolny equations (\ref{bog1}), (\ref{bog2}) 
for the solitons of gauged $O(3)$ sigma model.
Actually we can easily find the Bogomolny bound from the
supersymmetry algebra in Eq.~(\ref{alg}) by noting the fact that 
the anticommutator in  Eq.~(\ref{alg}) is hermitian, the
trace of its square is positive semi-definite, so that
we obtain the bound on the energy
\begin{eqnarray}
\sum_{A,B} \{Q^{(A)}_\alpha,Q^{(B)}{}^\beta\}
\{Q^{(A)}{}^\alpha,Q^{(B)}_\beta\}\ge 0  ~~{\rm or}~~ E \ge |Z|
\label{en}
\end{eqnarray}
in the frame where $P^i=0$.

The Bogomolny equations that saturate the energy bound are obtained
explicitly from the supersymmetry algebra (\ref{alg}). 
{}First, note the identity that follows from Eq.~(\ref{alg}):
\begin{eqnarray}
\label{ez}
E =\pm Z +\{ (\gamma^0 Q_{\pm})_\alpha, Q^\alpha_{\pm}\} ,
\end{eqnarray}
where 
\begin{eqnarray}
Q_{\pm}&=& Q^{(1)}\pm i \gamma^0 Q^{(2)}\nonumber \\
&=&\int d^2x \Big[
D_\nu\phi^a\gamma^\nu\gamma^0  \pm i
(\phi\times D_\nu\phi)^a \gamma^0\gamma^\nu\gamma^0
-i(n^a\gamma^0 \mp i(n\times\phi)^a ) f'
\Big]\psi^a.
\label{qq}
\end{eqnarray}
It is clear that Eq.~(\ref{ez}) is precisely 
saturated for those field configurations $|\alpha>$ such that
$Q_{+}|\alpha>=0$ or $Q_{-}|\alpha>=0$. 
Equivalently, we obtain the condition  
\begin{eqnarray}
\label{sbogo}
D_\nu\phi^a\gamma^\nu\gamma^0  \pm i
(\phi\times D_\nu\phi)^a \gamma^0\gamma^\nu\gamma^0
-i(n^a\gamma^0 \mp i(n\times\phi)^a ) f'=0.
\end{eqnarray}
Of course, this condition is equivalent to 
the Eqs.~(\ref{bog1}) and (\ref{bog2}).
Taking the trace of this equation  with the Gauss law constraint 
gives Eq.~(\ref{bog2}), and  multiplying by $\gamma^i$ and taking the
trace yields Eq.~(\ref{bog1}). 
We can also obtain the same equation (\ref{sbogo}) 
by letting  the supersymmetry
transformation of $\psi^a$ field generated by $Q_\pm$ vanish.
We have thus derived the Bogomolny relations
from the requirement of $N=2$ supersymmetry for the gauged $O(3)$ sigma 
model. 

Notice that a nontrivial soliton configuration satisfying
the Bogomolny equations breaks a half of supersymmetry, 
either the one generated by $Q_{+}$ or the one by $Q_{-}$.
This is a generic feature of 
all self-dual models with extended supersymmetry~\cite{witten78}.

In summary, we have constructed the $N=2$ supersymmetric gauged $O(3)$
sigma model with a Chern-Simons term and studied the relation
between the Bogomolny relations and the $N=2$ extended supersymmetry algebra.
We have shown that finding a conserved charge within the $N=1$
supersymmetric model we could extract another supersymmetry of the model.

It would be interesting to find out the similar construction of
$N=2$ supersymmetry in the gauged $CP^N$ model 
with Chern-Simons term as well as 
this model with Maxwell term.

\vskip 2em
\noindent{\large\bf Acknowledgments}\\
We would like to thank Kimyeong Lee for critical comments and
careful reading of this manuscript. This work was supported
by the Korea Science and Engineering Foundation(KOSEF) through
the Center for Theoretical Physics at Seoul National University.
\end{document}